\documentclass[sigconf,nonacm]{acmart}            
\AtBeginDocument{%
  }

\setcopyright{acmlicensed}
\copyrightyear{2025}
\acmYear{2025}
\acmDOI{XXXXXXX.XXXXXXX}
\acmConference[Conference acronym 'XX]{Make sure to enter the correct
  conference title from your rights confirmation email}{June 03--05,
  2018}{xxxx, xxxx}
\acmISBN{978-1-4503-XXXX-X/2026/04}

\usepackage{amssymb}

\usepackage[edges]{forest}
\usepackage{xcolor}
\usepackage{tcolorbox}
\usepackage{pifont}

\definecolor{taskcolor}{RGB}{63, 81, 181}
\definecolor{designcolor}{RGB}{0, 150, 136}
\definecolor{qualitycolor}{RGB}{255, 152, 0}
\definecolor{envcolor}{RGB}{156, 39, 176}
\definecolor{llmcolor}{RGB}{233, 30, 99}
\definecolor{outputcolor}{RGB}{33, 150, 243}

\begin{document}



\title{Designing Empirical Studies on LLM-Based Code Generation: Towards a Reference Framework}

\author{Nathalia Nascimento}
\email{nqm5742@psu.edu}
\affiliation{%
  \institution{The Pennsylvania State University}
  \city{Great Valley, Malvern}
  \state{PA}
  \country{USA}
}

\author{Everton Guimaraes}
\email{ezt157@psu.edu}
\affiliation{%
  \institution{The Pennsylvania State University}
  \city{Great Valley, Malvern}
  \state{PA}
  \country{USA}
}

\author{Paulo Alencar}
\email{palencar@uwaterloo.ca}
\affiliation{%
  \institution{University of Waterloo}
  \city{Waterloo}
  \country{Canada}}

\renewcommand{\shortauthors}{Nascimento et al.}

\begin{abstract}
The rise of large language models (LLMs) has introduced transformative potential in automated code generation, addressing a wide range of software engineering challenges. However, empirical evaluation of LLM-based code generation lacks standardization, with studies varying widely in goals, tasks, and metrics, which limits comparability and reproducibility. In this paper, we propose a theoretical framework for designing and reporting empirical studies on LLM-based code generation. The framework is grounded in both our prior experience conducting such experiments and a comparative analysis of key similarities and differences among recent studies. It organizes evaluation around core components such as problem sources, quality attributes, and metrics, supporting structured and systematic experimentation. We demonstrate its applicability through representative case mappings and identify opportunities for refinement. Looking forward, we plan to evolve the framework into a more robust and mature tool for standardizing LLM evaluation across software engineering contexts.
\end{abstract}




\maketitle

\section{Introduction}

Large Language Models (LLMs) are rapidly transforming software engineering, particularly in the area of automated code generation \cite{gu2024effectiveness}. While recent research has demonstrated the potential of LLMs to generate functional code, the empirical evaluation of these models remains largely fragmented. Studies often adopt ad hoc experimental setups, resulting in limited reproducibility, poor comparability, and challenges in generalizing findings across tasks, models, and contexts. As Baltes et al. argue \cite{baltes2025guidelines}, while empirical research in software engineering is supported by well-established guidelines, LLMs introduce unique challenges—such as non-determinism, evolving versions, and limited transparency—that demand tailored methodologies to ensure validity and reproducibility. This gap highlights the need for domain-specific frameworks to guide empirical investigation in LLM-based code generation.

Empirical experimentation is a foundational practice in software engineering research \cite{wohlin2012experimentation}, enabling rigorous assessment of techniques against well-defined criteria. However, within LLM-based code generation, there is no standard methodology to guide how experiments should be designed, executed, or interpreted. Studies vary widely in problem sources, evaluation goals, metrics, and environmental conditions—making it difficult to build cumulative knowledge or derive best practices.


This paper proposes a bottom-up framework for empirical investigation in LLM-based code generation. Rather than prescribing a fixed methodology, our approach distills common patterns and recurring elements from existing empirical studies to synthesize a generalizable structure. The framework identifies and organizes core components of empirical design—such as problem sources, quality attributes, and evaluation metrics—while also exposing variability points that define open research opportunities.

By formalizing these elements and their interrelationships, the framework promotes consistency, comparability, and reproducibility in future experiments. It aims at supporting researchers in designing better-grounded studies, systematically measuring key variables such as correctness, efficiency, and bias, and uncovering underexplored dimensions of LLM behavior in software engineering contexts.


\section{Related Work}

Several recent efforts have proposed frameworks to support empirical research involving large language models (LLMs) in software engineering, particularly code generation.

Schneider et al.~\cite{schneider2025reference} introduce a reference model for empirically comparing LLMs with humans, emphasizing decision points and dependencies in experimental setups. While they address fairness in human-versus-LLM evaluations, our framework generalizes beyond that scope, offering modular components for designing LLM experiments regardless of human baselines.

Yeo et al. \cite{yeo2024framework} propose a structured evaluation framework for assessing the code generation ability of large language models across different programming tasks. Their framework introduces a taxonomy of task categories, input-output formats, and evaluation metrics, with a focus on capturing both functional and non-functional properties such as correctness, performance, and robustness. While Yeo et al. concentrate on categorizing what should be evaluated and how, our work emphasizes how experiments themselves should be constructed.

De Martino et al.~\cite{de2024framework} propose PRIMES, a framework tailored to LLM-based software repository mining. Based on insights from two empirical studies, it offers practical guidance for prompt engineering and data extraction. 
Wagner et al. \cite{wagner2025towards} present the first holistic set of guidelines for empirical studies involving LLMs in software engineering. Their work classifies different study types (e.g., using LLMs as annotators, judges, or study subjects) and proposes preliminary best practices to improve reproducibility and reporting quality. In contrast to their focus on guidelines and study-type classification, our approach provides a structured, bottom-up framework that identifies and organizes the core elements of empirical design and highlights variability points to support the generation of new experiment instances.

\section{Research Method}
\label{sec:method}

To guide the development of our proposed framework, we conducted a structured search in the ACM Digital Library targeting empirical studies involving LLMs in code generation tasks. Using the following boolean query, each term was searched in both the \texttt{title} and \texttt{abstract} fields:

\begin{tcolorbox}[colback=gray!5!white, colframe=gray!75!black, title=Search String]
\small
((LLM OR LLMs OR "large language model" OR "large language models" OR ChatGPT)
AND
("code generation" OR "program synthesis" OR coding OR programming)
AND
(empirical AND (compar* OR evaluation OR study OR experiment)))
\end{tcolorbox}

This query retrieved 75 papers published between 2023 and 2025. After applying inclusion and exclusion criteria, 32 papers were retained, focusing on empirical evaluations of LLMs in code generation tasks. For this study, we selected the 11 most cited papers from the dataset and 2 additional papers identified via snowballing: \cite{paper1,paper2,paper3,paper4,paper5,paper6,paper7,paper8,paper9,paper10,paper11,gu2024effectiveness,nascimento2025effective}. Of these, 9 papers informed the construction of the framework by revealing recurring experimental patterns, while 2 were used to evaluate how the framework generalizes to previously unseen setups. The full dataset, including selection justifications, is publicly available at \cite{anonymous_2025_17230476}.

\textbf{Inclusion/Exclusion Criteria:} We included papers presenting empirical evaluations of LLMs on code generation tasks, especially those introducing or applying benchmarks, metrics, or experimental designs. We excluded studies focused solely on education, user perception, or tasks unrelated to code generation (e.g., translation or bug repair), as well as non-empirical position or vision papers. 

\section{Framework Grounding and Design} \label{sec:framework}

\begin{figure}[!htb]
\centering

\begin{forest}
for tree={
    font=\ttfamily\scriptsize,
    grow'=0,
    child anchor=west,
    parent anchor=south,
    anchor=west,
    calign=first,
    inner sep=1pt,
    s sep=1pt,
    l sep=10pt,
    edge path={
        \noexpand\path [draw, \forestoption{edge}]
        (!u.south west) +(5pt,0) |- (.west)\forestoption{edge label};
    },
    before typesetting nodes={
        if n=1
            {insert before={[, phantom]}}
            {}
    },
    fit=band,
    before computing xy={l=15pt},
}
[Framework
  [1. Coding Task
    [1.1. Description \ding{51} (generate programming solutions)]
    [1.2. Application Context
      [Software Engineering \ding{51} (competitive programming)]
      [Data Science]
      [Web Application]
      [Internet of Things]
      [Machine Learning]
    ]
    [1.3. Problem Source
      [1.3.1. Data Sources
        [GitHub]
        [LeetCode \ding{51}]
        [Kaggle]
        [HumanEval Dataset]
        [Stack Overflow]
      ]
      [1.3.2. Application Scenarios
        [Robotics Tasks]
        [Unsupervised Learning]
      ]
    ]
  ]
  [2. Quality and Metrics Evaluation
    [2.1. Quality Attribute
      [Correctness \ding{51}]
      [Time Efficiency \ding{51} (runtime execution)]
      [Energy Efficiency]
      [Bias Assessment]
      [Maintainability]
      [Security]
      [Code Complexity]
      [Readability]
    ]
    [2.2. Metric
      [Success Rate \ding{51} (pass/fail of coding tasks)]
      [Pass@1]
      [Execution Time \ding{51}]
      [Cyclomatic Complexity]
      [Memory Consumption \ding{51}]
      [Energy Consumption]
      [Cosine Code Similarity]
      [Maintainability Index]
      [Code Security Metric]
      [Subjective Evaluation [Open-ended Feedback|Behavioral Observation|Ratings]]
    ]
  ]
  [3. Empirical Research
    [3.1. Method
        [Controlled Experiment \ding{51}]
        [Survey/Questionnaire]
        [Case Study]
    ]
    [3.2. Hypothesis and Experimental Design
      [3.2.1 Goal-Question-Metrics]
      [3.2.2 Hypothesis Formulation \ding{51}]
      [3.2.3 Baseline Selection \ding{51}]
      [3.2.4 Variables \ding{51}]
    ]
    [3.4. Data Analysis
        [Quantitative: Hypothesis Testing \ding{51}|Descriptive Stats \ding{51}]
        [Qualitative: Manual Review|Expert Feedback|Developer Perception]
    ]
    [3.4. Comparative Setup
        [LLM vs. LLM]
        [LLM vs. Human \ding{51} (novice and expert programmers)]
        [LLM vs. ML-based System]
    ]
  ]
  [4. Environment
    [4.1. Computational Resources \ding{51} (execution environment for LLM)]
    [4.2. Hardware Constraints]
  ]
  [5. LLM Model
    [5.1. Prompt Engineering]
    [5.2. Parameter Tuning]
    [5.3. Model Selection \ding{51} (ChatGPT)]
  ]
  [6. Generated Code/Output
    [Code Snippets \ding{51}]
    [Task Solutions \ding{51}]
    [Generated Behaviors]
  ]
]
\end{forest}

\caption{Overview of the proposed framework instantiated based on the study~\cite{paper8}.}
\label{fig:framework-tree}
\end{figure}
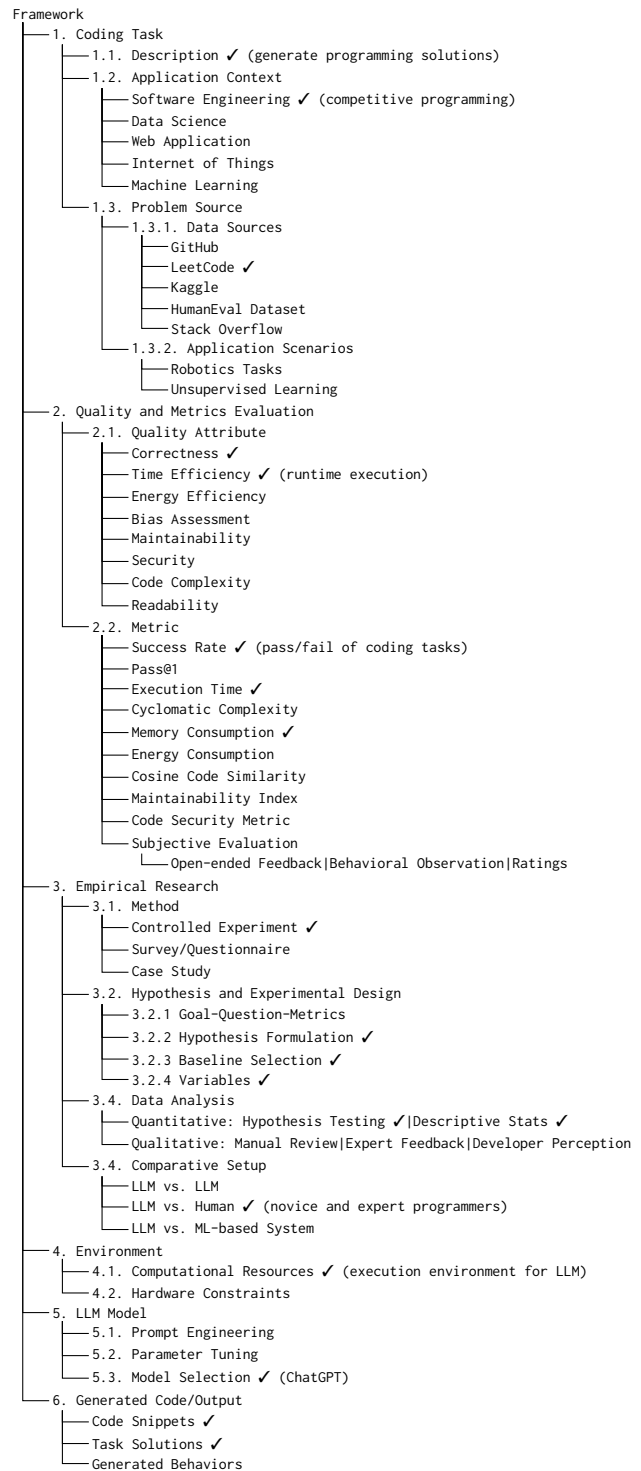

The proposed framework was developed using a bottom-up approach, grounded both in our own experience conducting empirical research on LLM-based software engineering (\cite{paper8,nascimento2023gptloop,nascimento2025effective,guimaraes2025}) and in the analysis of selected empirical studies from the literature. 

Figure~\ref{fig:framework-tree} presents the framework structure, organized into six core components that reflect key elements of LLM-based code generation experiments: \textbf{(1) Coding Task}, \textbf{(2) Quality and Metrics Evaluation}, \textbf{(3) Empirical Research}, \textbf{(4) Environment}, \textbf{(5) LLM Model}, and \textbf{(6) Generated Output}.

Each component defines a configurable space that can be instantiated based on specific experimental goals and contexts. For example, a study may define \textit{Quality Attributes} such as \texttt{Correctness} and \texttt{Energy Efficiency}, guiding the selection of corresponding evaluation metrics—like \texttt{Pass@1} or \texttt{Energy Consumption}. Similarly, the \textit{Coding Task} component may utilize problem sources such as LeetCode or GitHub, while the \textit{Empirical Research} branch may define a controlled experiment comparing two LLMs or benchmarking against human performance.

To illustrate its applicability, Figure~\ref{fig:framework-tree} also shows how an existing empirical study can be instantiated from this structure. Specifically, we highlight the components used in the study ``Artificial Intelligence vs. Software Engineers: An Empirical Study on Performance and Efficiency using ChatGPT" \cite{paper8}, which evaluates ChatGPT's performance and efficiency on LeetCode-style programming tasks, benchmarking it against novice and expert human programmers. The selected components are marked with \ding{51}, covering aspects such as correctness, execution time, memory usage, and controlled experimental setup.

The modular design of the framework promotes consistency and traceability in LLM evaluation studies while remaining flexible enough to accommodate novel experimental designs and emerging research priorities.

\section{Overview of Framework Components}

As mentioned, the framework components were also grounded by frequently adopted practices in LLM-based code generation studies (e.g., \cite{10.1145/3697012,paper2,paper3,paper5,paper6,paper7}) and structured to reflect common experimental design choices. While Figure~\ref{fig:framework-tree} highlights prevalent instantiations, the framework remains extensible.

\subsection{Problem Sources}
\begin{itemize}
  \item \textbf{GitHub}: Open-ended, real-world scenarios \cite{paper7}.
  \item \textbf{LeetCode / APPS}: Algorithmic and competitive problems \cite{paper5,paper8}.
\end{itemize}

\subsection{Application Context}
LLM performance varies across domains. Gu et al.~\cite{10.1145/3697012}, Rasnayaka et al.~\cite{paper2}, and Nascimento et al.~\cite{nascimento2025effective} show how domain-specific settings (e.g., web, data science) affect code generation outcomes.

\subsection{Quality Attributes}
Drawing from ISO/IEC 25010~\cite{ISO25010} and empirical literature, we group quality concerns into:
\begin{itemize}
  \item \textbf{Functional Quality}: Correctness and completeness \cite{paper1,paper5}.
  \item \textbf{Technical Quality}: Readability, modularity, complexity \cite{paper5}.
  \item \textbf{Resource Efficiency}: Runtime, memory, and energy usage \cite{paper3,paper8}.
  \item \textbf{Ethical/Social Quality}: Fairness, bias, and security risks \cite{paper6,paper7,paper10,paper11}.
\end{itemize}

\subsection{Evaluation Metrics}
Metrics are selected to quantify quality attributes. Examples include:
\begin{itemize}
  \item \textbf{Correctness}: Test pass rate, compilation success \cite{paper1,paper5}.
  \item \textbf{Complexity}: Cyclomatic complexity, token count \cite{paper3}.
  \item \textbf{Security and Bias}: CWE violations \cite{paper7}, fairness audits \cite{paper6}, and adversarial prompts \cite{paper11}.
  \item \textbf{Efficiency}: Execution time and memory profiling \cite{paper3,paper8}.
\end{itemize}

\section{Potential Instances of the Framework}

To assess the applicability and extensibility of our proposed framework, we randomly selected \textbf{two representative studies} from the included dataset. Table~\ref{tab:framework-instances} summarizes how each study maps to the core components of our framework, along with potential extensions inspired by their specific research goals and designs. 

\begin{table*}[t]
\centering
\caption{Mapping of Framework Components to Empirical Instances (papers \cite{paper1} and \cite{paper4}) and Suggested Extensions}
\label{tab:framework-instances}
\begin{tabular}{|p{3.3cm}|p{6.1cm}|p{6.1cm}|}
\hline
\textbf{Framework Component} & \textbf{Instance 1: Ouyang et al.~\cite{paper1}} & \textbf{Instance 2: Ren et al.~\cite{paper4}} \\
\hline
\textbf{Coding Task} & Python-based algorithmic problems focusing on reproducibility and stability & Java-based exception-handling tasks extracted from documentation \\
\hline
\textbf{Problem Source} & HumanEval, CodeContests, and APPS public benchmarks & 3,079 tasks derived from official Java API documentation \\
\hline
\textbf{Quality Attributes} & Correctness and Consistency (output stability under repeated generations) & Correctness, Maintainability, and Security \\
\hline
\textbf{Evaluation Metrics} & pass@k, semantic/syntactic similarity, structural comparisons, dispersion measures & Bug count, specification adherence (static), runtime bug reduction (dynamic) \\
\hline
\textbf{Empirical Method} & Controlled experiment with multiple generations per prompt (varying temperature) and statistical comparison & Controlled comparison of prompt variants, evaluating effectiveness of prompt chaining strategies \\
\hline
\textbf{LLM Configuration} & ChatGPT with temperature and sampling parameter variation & ChatGPT with knowledge-driven prompt chaining (KPC) and fine-grained prompt segmentation \\
\hline
\textbf{Extension Opportunities} 
& \begin{itemize}
  \item Introduce \texttt{Stability} as a new Quality Attribute
  \item Add output variance/dispersion to Evaluation Metrics
  \item Explicitly model temperature and sampling parameters in LLM configuration
\end{itemize}
& \begin{itemize}
  \item Define ``Exception Handling" as a specialized task category
  \item Model Prompt Chaining under Prompt Engineering
  \item Introduce Specification Conformance as a metric
\end{itemize}
\\
\hline
\end{tabular}
\end{table*}

\subsection{Instance 1: Capturing Non-Determinism in Code Generation}

Ouyang et al.~\cite{paper1} evaluate ChatGPT’s non-determinism across three public code-generation benchmarks. Their work illustrates how our framework supports studies assessing model stability and reproducibility. It also highlights gaps: the need to formalize \textit{stability} as a quality attribute, to include variance-based metrics for output variability, and to make sampling parameters (e.g., temperature) explicit under model configuration.

\subsection{Instance 2: Enhancing Exception Handling via Prompt Chaining}

Ren et al.~\cite{paper4} investigate a prompt-chaining method (KPC) to improve exception-handling code generation. This demonstrates the framework’s adaptability to specialized tasks and advanced prompting strategies. It also points to extensions: defining exception handling as a specific task category, explicitly modeling chaining strategies under prompt engineering, and formalizing metrics like specification adherence and runtime bug reduction.

\section{Conclusion}

This paper introduces a theoretical framework for designing empirical experiments in LLM-based code generation, developed through a bottom-up process grounded in both our own experimental experience and a review analysis of recent literature. The framework is structured around key components such as problem sources, quality attributes, and evaluation metrics, supporting reproducibility, comparability, and coverage across diverse experimental setups. 

We illustrated its applicability by mapping two representative studies not used in the construction of the framework, which revealed additional elements—such as non-determinism analysis and prompt chaining strategies—that can be formally integrated. This illustrates not only the coverage and adaptability of the framework, but also its maturity as a living artifact that evolves alongside the research landscape. As the framework continues to mature, it aims to support more standardized and comprehensive experimentation in LLM-based software engineering. 

\section{Future Plans}

\textbf{Literature Review to Guide Framework Refinement:}  
Our proposed framework has been informed by a preliminary search and analysis of selected papers from the dataset reported in \cite{anonymous_2025_17230476}, combined with our practical experience in the field. As part of our future work, we will systematically evaluate these papers to construct a traceability matrix linking each paper in the dataset to the elements of our proposed framework. This matrix will help identify underexplored domains, frequently used problem sources, overlooked quality attributes, and applied evaluation metrics, ultimately guiding the refinement and extension of the framework.


\textbf{Expansion of Framework Components:}  
In addition to extending the existing list of components, we plan to explore the inclusion of new dimensions, such as:  
\emph{Reproducibility Factors} (e.g., seed control, open datasets, model versioning);  
\emph{Data Collection Strategies} (e.g., execution logs, generated code, prompts, developer feedback); and  
\emph{Evaluation Strategies} (e.g., automated evaluation, human-in-the-loop assessment, peer review, and expert validation).

\textbf{Design of Novel Experiment Instances:}  
The proposed framework will be applied to design new empirical studies on LLM-based code generation, particularly focusing on gaps identified in the literature. These novel experimental instances will demonstrate how the framework can be adapted to diverse contexts and research goals.

\textbf{Automatic Design of Research Protocols:}  
We envision this framework evolving into an interactive tool for supporting the design of controlled experiments. Researchers will be able to specify the application domain (e.g., mobile apps, robotics, healthcare) and define a Goal-Question-Metric (GQM) strategy. The tool will then recommend research questions, quality attributes, and evaluation metrics aligned with the chosen goals. As output, a complete research protocol will be generated, including the GQM, hypotheses, experimental design, subjects, evaluation methods, and guidance for execution.

\textbf{Curated Dataset of Empirical Research:}  
To support protocol recommendations, we will maintain a database of existing empirical studies in LLM-based code generation. This dataset will guide researchers toward addressing less-explored aspects of the literature and foster diversity in experimental design.

\textbf{Automation of Controlled Experiment Design and Execution:}  
Beyond research protocol generation, we aim to automate parts of the experimental pipeline. For instance, consider an experiment comparing different LLMs on mobile app development tasks. A software agent could automatically scrape problems from relevant sources, use various LLMs to solve them, generate datasets with the results, and perform statistical analyses. This would streamline experimental workflows and enhance reproducibility.

\textbf{Extension to Other Software Engineering Tasks:}  
Although the current framework focuses on code generation, it is inherently adaptable and can be extended to a wider range of software engineering tasks. Future work will explore its application in empirical investigations involving the use of LLMs for tasks such as unit test generation, requirements elicitation, bug fixing, documentation synthesis, and code refactoring. These domains introduce new problem sources, quality attributes, and evaluation metrics, which will be progressively incorporated into the framework.

\bibliographystyle{ACM-Reference-Format}
\bibliography{references}

\end{document}